\documentclass{article}
\pdfoutput=1

\PassOptionsToPackage{numbers, compress}{natbib}
\usepackage[preprint]{neurips_2019}
\usepackage[utf8]{inputenc} % allow utf-8 input
\usepackage[T1]{fontenc}    % use 8-bit T1 fonts
\usepackage{hyperref}       % hyperlinks
\usepackage{url}            % simple URL typesetting
\usepackage{booktabs}       % professional-quality tables
\usepackage{amsfonts}       % blackboard math symbols
\usepackage{nicefrac}       % compact symbols for 1/2, etc.
\usepackage{microtype}      % microtypography
\usepackage{amsmath}
\usepackage{amssymb}
\usepackage{graphicx}

\title{Fair treatment allocations in social networks}

\author{%
  James Atwood \\
  Google Research \\
  \texttt{atwoodj@google.com} \\
  \And
  Hansa Srinivasan \\
  Google Research \\
  \texttt{hansas@google.com} \\
  \And
  Yoni Halpern \\
  Google Research \\
  \texttt{yhalpern@google.com} \\
  \And
  D. Sculley \\
  Google Research \\
  \texttt{dsculley@google.com} \\
}

\begin{document}

\maketitle

\begin{abstract}
Simulations of infectious disease spread have long been used to understand how epidemics evolve and how to effectively treat them.  However, comparatively little attention has been paid to understanding the fairness implications of different treatment strategies – that is, how might such strategies distribute the expected disease burden differentially across various subgroups or communities in the population?  In this work, we define the \emph{precision disease control problem} -- the problem of optimally allocating vaccines in a social network in a step-by-step fashion -- and we use the ML Fairness Gym~\citep{fairnessgym} to simulate epidemic control and study it from both an efficiency and fairness perspective.  We then present an exploratory analysis of several different environments and discuss the fairness implications of different treatment strategies.
\end{abstract}

\section{Introduction}

The field of public health has long understood that different medical and health problems are not always easily decomposable into individual health measures, but can benefit from a networked analysis - viewing health as a population-level process.
Examples of this type of analysis have long been applied to spread of infectious diseases, and more recently to other health factors like obesity~\citep{christakis2007spread}, substance abuse~\citep{rosenquist2010spread}, happiness~\citep{fowler2008dynamic}, and even misinformation about health topics~\citep{fernandez2015health}.

In network analyses, determining who benefits from an intervention is not as simple as observing who is directly affected by the intervention (e.g., who receives a vaccine), since neighbors and neighbors of neighbors are also affected indirectly.  With this in mind, we can still investigate how different health interventions differentially benefit individuals and communities within a larger population. These tradeoffs are important to study because they highlight how pursuing a coarse metric of success of an intervention averaged over an entire population (like e.g., minimizing total expected number of sick-days) can, under certain circumstances, leave parts of the population under-served.  

We study a stylized version of the public health task of epidemic control that highlights the networked setting. More specifically, we study the problem of optimally allocating vaccines within a social network in a step-by-step manner as the disease spreads.  We assume that the agents allocating vaccines can observe the disease states of individuals within the population and the underlying contact network structure.  We call this the {\em precision disease contagion control} problem.

This work is relevant to fairness in machine learning for healthcare because it addresses class of optimization problems that appear repeatedly in the public health. We study this problem in simulation in order to illuminate core dynamics \cite{epstein2008} -- with the rise of learned policies that attempt to optimize health outcomes on real data, the tensions at the root of the underlying optimization problem become increasingly relevant.

We propose this environment as a tool for researchers to explore the dynamics of disease under different contact and initial infection conditions, and to evaluate the relative performance of allocation approaches, learned or otherwise, that they may be interested in exploring.

This work is exploratory in nature.  We present several different contexts in which epidemics can spread and explore the efficacy and fairness of different treatment scenarios within those contexts.

\subsection{Contributions}
\begin{itemize}
    \item We pose a version of the precision disease contagion control problem and discuss heuristic and learned policies. 
    \item We characterize some of the tradeoffs between efficiency and fairness in simple social network graphs.
    \item We open-source all of our code in an extensible manner to provide for reproduction and extensions of this work.
\end{itemize}

\section{Background and Related Work}

\begin{figure}
    \centering
    \includegraphics[width=\textwidth]{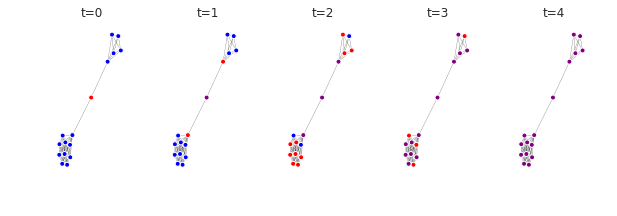}
    \caption{An illustration of the development of an epidemic on a barbell graph with a central individual.  Here, blue nodes are in the `Susceptible' state, red nodes are in the `Infected' state, and purple nodes are in the `Recovered' state.  `Recovered` is an abosrbing state. After t=4, there are not more transitions to take place.}
    \label{fig:barbell_evolution}
\end{figure}

Mathematical models for studying infectious disease spread have a long history~\citep[e.g.,][]{longini1978optimization, hethcote1978immunization, patel2005finding, tennenbaum2008simple}, including explicit treatment of network structures~\citep{newman2002spread, valente2012network, newman2018networks}, and evolving control policies~\citep{sharomi2017optimal}. 

\citet{keeling2012optimal} used analytical models to illustrate a tension between optimal and equitable distributions of vaccines in a simple case of two non-interacting (or barely interacting) populations. \citet{yi2015fairness} study tradeoffs between fairness and effectiveness when considering different possible prioritization orderings of who receives vaccines in the case of limited resources. \citet{salathe2010dynamics} discuss the importance of ``intercommunity-bridges'', individuals who have contact with multiple communities, in order to stem the spread of disease.

Ethical discussions of prioritizing individuals for treatment based on considerations like quality-adjusted life years, vulnerability, need, productivity, ability to treat others, and lottery have been discussed extensively~\citep[e.g.,][]{emanuel2006should, persad2009principles, buccieri2013ethical, saunders2018equality}, but generally deal with the problem of a general order of prioritization rather than precision treatments and do not consider explicit social network structures.

To the best of our knowledge, the problem of optimally allocating vaccines within a social network in a step-by-step manner as the disease spreads has not been explicitly studied neither from an effectiveness nor fairness view. The turn-based precision version of the problem is closer to learning to play a strategy game like chess or go which has received a lot of attention in the reinforcement learning community~\citep[e.g.,][]{atari, silver2017mastering}. 

\subsection{Fairness in networks} 
\label{sec:fairness_in_networks}

Many recently proposed measurements of fairness focus either on group fairness~\citep{hardt2016equality} or  individual-fairness~\citep{dwork2012fairness}. Networks serve as an interesting setting because of they way that they can richly reproduce many societal structures.

{\bf Communities:} Social networks tend to organize in community structures~\citep{girvan2002community, faust2005using}. Membership in these communities tends to be much more nuanced than strictly binary or categorical with individuals potentially belonging to many communities, possibly with differing degrees of identification~\citep{palla2005uncovering} and with communities interacting at different scales or resolutions~\citep{ronhovde2009multiresolution}. Using measures like overall burden to a community rather than probability of disease conditioned on being from a community, directly links fairness measures to group well-being. 

{\bf Centrality:} Individuals in different positions in the graph may play different roles and be treated differently. Centrality of a node in a network refers to a measure of its influence on the network and can be characterized using a number of different measures~\citep{lu2016vital}. Three important measures are {\em degree centrality}: the number of immediate neighbors; {\em betweenness-centrality}: the number of shortest paths that pass through this node; and {\em eigenvector centrality}: uses the dominant eigenvector of the adjacency matrix.

Even with the same set of actors, an individual's characterization in the network (both in terms of community and centrality) may be very different depending on how the edges in the network are defined. For example, in an workplace, the communities formed by who eats lunch together may be different from the communities of who attend meetings together. Thus centrality and community memberships of an individual should not be thought of as unique but depend crucially on the context being analyzed.

\section{Formalizing the precision contagion control problem}
\begin{figure}
    \centering
    \includegraphics[width=0.9\columnwidth]{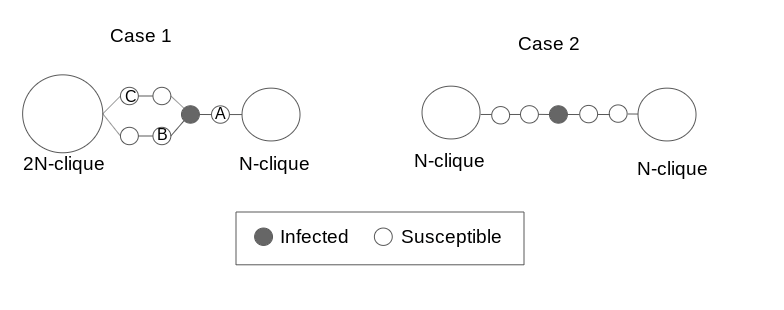}
    \caption{Network structures to illustrate some of the complexity of the precision control task. {\bf Case 1 - greedy suboptimal:} If one treatment is available per time-step, a greedy agent would prefer to treat node A whereas an agent that plans ahead would treat B then C. {\bf Case 2 - wait and see:} If disease transfer occurs with with $\tau = 0.5$ and only one treatment is available overall, an agent minimizing expected sick days should to wait to see if the disease transfers to each neighbor before deciding which side to treat.}
    \label{fig:greedy_suboptimal}
\end{figure}
We pose the precision contagion control problem in this section. In defining the problem, we build on work on discrete-time compartmental modeling of infectios disease \cite{brauer2010discrete,hernandez2015discrete,liu2015effect,allen1994some}.

A population of individuals, $V$, is connected by edges, $E$, forming a social network $N = (V, E)$. Individuals can be in one of three health states $\{S, I , R\}$ for susceptible, infected, recovered. The health of the network is $H \in \{S, I, R\}^{|V|}$. Health evolves in discrete time steps. At time $t_0$ an initial set of nodes $V_0$ are infected. At every subsequent step step, disease spreads stochastically from infected to susceptible neighbors at every time step, with probability:

\begin{equation}
p_{S \rightarrow I} = 1-(1-\tau)^{\#I(N)},
\end{equation}
where $\tau \in [0, 1]$ is probability of transmission and $\#I(N)$ is a count of infected neighbors.

The recovery process does not depend on neighborhood.
\begin{equation}
p_{I \rightarrow R} = \rho,
\end{equation}

where $\rho \in [0, 1]$ is probability of recovery. 

An agent allocates $N_t$ treatments at each step. Treatments effect a direct transition to the R state, and have no effect on individuals in states $\{I, R\}$ (i.e., treatments are {\em preventative}). Figure~\ref{fig:barbell_evolution} illustrates the disease being transmitted and individuals recovering in a simple network with no treatments bring allocated.

The precision contagion control problem consists of learning a treatment policy $\pi(N, \tau, \rho, N_t):  H \rightarrow (V \cup \{\varnothing\})^{N_t}$, i.e., finding the best next individuals to treat based on the current health state of the network. There is always an option to not use an available treatment which we denote by allocating that treatment a non-existent individual, $\varnothing$.

In public health, uncertainties in the network structure as well as the logistical difficulties in implementing precision policies make it practically not worthwhile to optimize precision interventions in the way that we describe here. In this work, we consider fairness implications of this idealized precision setting that may become possible as health IT infrastructure makes it possible to target health interventions on the individual level. 

Finding optimal strategies for precision contagion control is not easy. The following properties of the problem, which we prove with example network structures, highlight some of the complexity of the task. 

{\bf Greedy is not optimal:}  A greedy policy that aims to minimize expected total disease burden at every step in a greedy way can suffer from being sub-optimal (Figure~\ref{fig:greedy_suboptimal}, Case 1).

{\bf Act now or wait?:} Strategies that wait to see how the disease progresses for some number of steps before allocating a treatment can be more effective than treating immediately (Figure~\ref{fig:greedy_suboptimal}, Case 2). 

From a fairness perspective, one might be interested in finding treatment strategies that equalize outcomes in the population.

{\bf Uniform allocation may not produce uniform outcomes:} Some individuals are more or less at risk before intervention due to their position in the social network. Uniformly allocating treatment does not always create uniform outcomes. This will be explored in more detail in Section~\ref{sec:chain}.

% \input{formal_problem}

% When thinking about how costs of a disease outbreak are distributed across a population, sick-days are a simple measure of cost and since the process is stochastic, we can talk about the expected number of sick days for an individual. The effect of a policy is described by how it changes the expected number of sick days for each individual. In general, treatment strategies are expected to reduce the overall expected burden of an outbreak on a population.

% \subsection{Social welfare orderings and collective utility functions}

% We consider three common collective utility functions~\citep{endriss2018lecture}: The {\bf utilitarian} collective utility simply takes the arithmetic average of individual utilities. The {\bf egalitarian} collective utility measures the minimal individual utility and the {\bf elitist} collective utility measures the maximal individual utility.

% TODO: Use the word interference.

% \subsection{Group fairness}
% Notions of group fairness tend to focus on parity of various statistical measurements between groups. These notions - like demographic parity, equalized opportunity, etc. - are formulated for classification tasks where true and false positives and negatives are present, and thus differ in the specific statistical feature they desire to be equal. In the context of disease in social network structures, we do not have a relevant confusion matrix. Instead, we use an intuitive definition of group fairness most similar to demographic parity: equality of disease prevalence across groups.

\section{Experiments}

We implement the simulated dynamics of this environment using the ML Fairness Gym~\citep{fairnessgym}. Centrality is computed with the networkx package's centrality measures. The code for the environment, agents, and experiments will be published open-source and is intended to be contributed to the ML Fairness Gym collection of examples.

\subsection{Policies for contagion control}
There are many possible policies for contagion control. In these experiments we consider:

{\bf Random selection}: Treatments are uniformly at random allocated to individuals in the population. 

{\bf Direct optimization:} Some networks are simple enough that policies can be computed by direct optimization of expected number of sick days.

{\bf Central-first heuristic}: Central-first heuristics treat more {\em central} individuals before less-central individuals. Centrality is discussed in more detail in~\ref{sec:fairness_in_networks}.

{\bf Deep Q-Network}: In cases where we cannot directly optimize over policies because the graph, we train a Deep Q-Network~\citep{mnih2015human} with a single hidden layer using the Dopamine library~\citep{castro18dopamine} for reinforcement learning. Details of the training can be found in Appendix~\ref{app:training_dqn}.

\subsection{Barbell Graph}

We start by considering disease transmission in a barbell-like graph where a central, initially-infected node is connected to two cliques of different sizes through intermediaries.  Figure \ref{fig:barbell_evolution} shows the natural progression of disease without intervention.

The barbell setting was previously considered by~\citet{keeling2012optimal} as a clear example of conflict between minimizing disease in the total population, and treating communities equally.

If a single vaccine is available, the optimal policy is to treat on the side of the larger community to block the spread of disease. However, this policy is clearly unfair to the smaller community. If multiple vaccines are available and it is impossible to completely block the spread to one community, \citet{keeling2012optimal} showed that the optimal allocation will change depending on the amount of vaccine available, but that for many values it would strongly favor one community or the other.

\subsection{Chain graph}
\label{sec:chain}

\begin{figure}
    \centering
    \includegraphics[width=0.45\columnwidth]{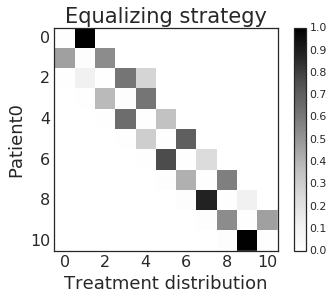}
    \includegraphics[width=0.45\columnwidth]{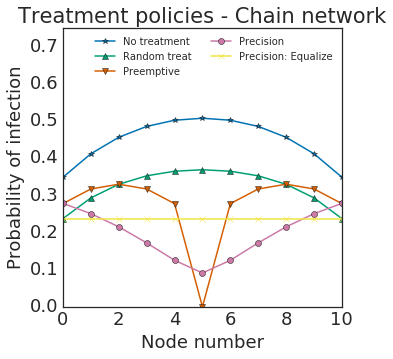}
    \caption{
    (Left) A visualization of the ``Precision: Equalize'' strategy. Each row represents a possible initial infected patient and the agent's conditional distribution of who to treat in response.
    (Right) Probability of infection of individuals in a chain graph using different treatment strategies. It is possible to ensure an equal probability of infection in the precision setting, but it does not have the lowest overall disease burden.}
    \label{fig:chain_treatments}
\end{figure}

In contrast to the barbell graph, the chain graph (where vertices are connected sequentially) has no clear group structure, but we can still consider likelihood of infection for individuals in different locations in the graph. For experiments we use a chain graph of size 11 with $\tau=0.75, \rho=1$ and a single treatment available.

If the initial infected patient is a uniformly chosen node in the network, the expected number of sick days with no intervention is higher for the central node and is lower at the two ends (Figure~\ref{fig:chain_treatments} Right - No treatment). A random allocation strategy recapitulates this shape. 

{\bf Preemptive treatment:} Treatment can be allocated preemptively, {\em before} the first patient is infected. The preemptive strategy that optimizes average utility is to treat the middle node in the chain. Interestingly, with only one treatment available, there may not be a preemptive strategy (including randomized strategies) that equalizes the expected disease burden for all individuals. We verify this for our chain graph by posing the problem as a constrained linear program (Appendix~\ref{app:chain_lp}) and determining that there is {\em no feasible solution}. This is in contrast to equalized-opportunity classifiers in the classification setting~\citep{hardt2016equality} which can always be constructed through randomization.

{\bf Precision treatment:} In the precision setting, we choose the individual to receive treatment {\em conditional} on the observation of which individual becomes sick first. In this setting, the strategy that optimizes average utility is to treat the individual beside the sick patient, choosing the side with more people. Finding a precision policy that equalizes disease burden can be formulated as a constrained linear program (Appendix~\ref{app:chain_lp}). Despite there being no preemptive treatment strategy that equalizes disease burden, a precision strategy {\em does exist} (Figure~ \ref{fig:chain_treatments} Left), though it has a higher average disease burden than the unconstrained policy. Figure \ref{fig:chain_treatments} compares how the different treatment policies distribute disease burden over the population in the chain graph.  The random policy is dominated by the precision equalizing strategy, i.e., every individual in the network fares equally or better with the precision equalize strategy than the random strategy. 

\subsection{Scale-free network}

\begin{figure}
    \centering
    \includegraphics[width=\textwidth]{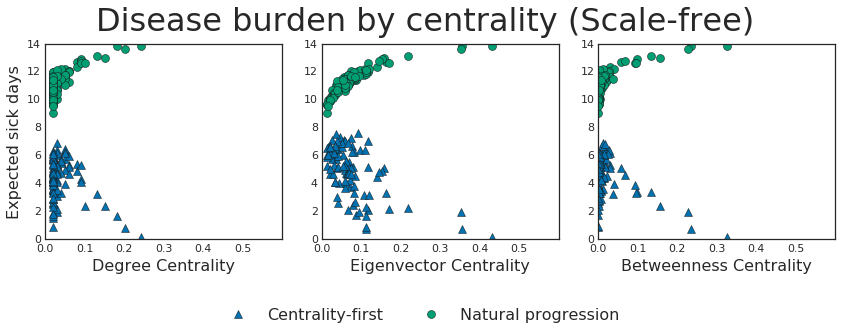}
    \caption{Disease burden by centrality (using three different measures of centrality) for the members of the synthetic scale-free network~\citep{holme2002growing}. Moderately central individuals fare the worst under centrality-based treatment.}
    \label{fig:general_graph}
\end{figure}

Many real social networks have degree distributions that follow a power-law (or similar) distribution, meaning that a small number of nodes have very high connectivity.

% Consider now a scale-free graph with a power-law degree distribution where uncommonly high levels of connectivity are found uncommonly often.  Within this context, we expect central nodes to play an outsize role in the propagation of an infection because of their position.

% How should vaccines be allocated in this scenario?  One approach is to distribute them randomly, which is superficially fair in the sense that everyone has a chance to avoid infection.  However, it is unclear how effective this strategy is in terms of overall utility (i.e. reduction in epidemic size).

% Another approach is to target central individuals.  We should expect this to outperform random allocation in terms of overall utility, but it is unclear how this plays out in terms of fairness.  Clearly, individuals who are more central in the contact graph will fare better than others because they receive vaccines.  

In this setting, how will individuals with low centrality fare if high-centrality individuals are prioritized for treatment? They may be better off than they would have been under random allocation if centrality-based allocation methods control the epidemic more effectively; they may be worse off because they do not receive treatment themselves.

We simulate this scenario with a synthetic scale-free network~\citep{holme2002growing} with 100 nodes and allowed an agent to allocate 1 treatment per turn. 
%For initial conditions, each individual draws from a Bernoulli with $p=0.25$ in order to determine whether they are infected. 
Infection spreads along contacts with $\tau=0.25$ and individuals recover with $\rho=0.01$.

Figure \ref{fig:general_graph} was generated by running 1000 simulations for 20 steps under three different centrality-based treatment allocation policies.  Note that moderately-central individuals have the highest epidemic size burden under centrality-based treatment because they do not enjoy the low risk that comes with being low centrality or the benefit offered to highly-central individuals by treatment.
%This raises questions of the fairness of the approach -- is an overall decrease in epidemic size worth a disproportionate burden being shouldered by individuals on the margin of the contact graph?

\subsection{Karate club graph}
\begin{figure}
    \centering
    \includegraphics[width=0.95\columnwidth]{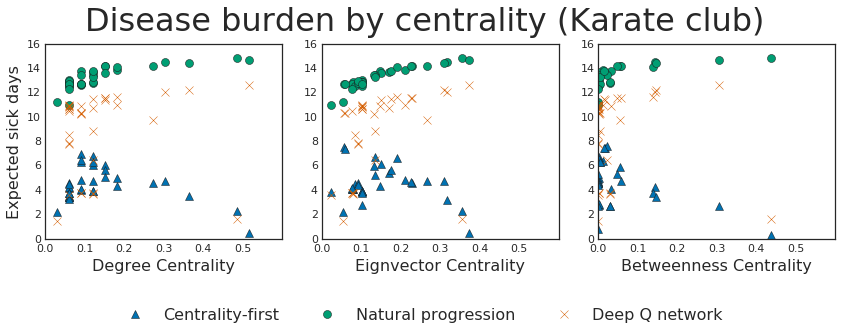}
    
    \caption{Disease burden by centrality (using three different measures of centrality) for each of the 34 members of the Karate club graph~\citep{zachary1977information}. With natural progression (green circles), more central nodes tend to have a higher expected number of sick days. Central-first treating heuristics bring down everyone's expected disease burden, with the highest disease burden for individuals with intermediate centrality. The Deep Q-network does not succeed in finding strategies that beat heuristics.}
    \label{fig:karate_club_centrality}
\end{figure}

\begin{figure}
    \centering
    \includegraphics[width=0.4\columnwidth]{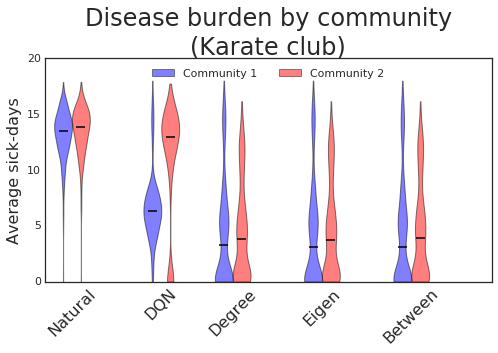}
    \caption{Violin plots showing disease burden in two communities in the karate club graph, for each of the treatment strategies. The distributions pictured are over 1000 simulation runs. Black lines indicate median runs of the simulation.}
    \label{fig:karate_club_community}
\end{figure}

Beyond the synthetic graphs discussed above, we also consider the small scale Karate club graph \citep{zachary1977information} which records the social interactions between 34 members of a karate club over three years, measuring the expected disease burden on individuals as a function of their centrality in the graph.

Simulations were run 1000 times with parameters ($\tau = \rho = 0.5; N_{t} = 1$). Each simulation lasts 20 time-steps. Agents prioritized by centrality (experiments were repeated with the same set of random seeds for each of the 3 different measures of centrality) or using a deep Q-network~\citep{atari}(Details in appendix~\ref{app:training_dqn}). Ties between individuals with equal centrality were broken randomly. In each run, the initial infected patient was chosen uniformly. 

With the natural disease progression, more central individuals tend to be {\em more at risk}, but once central-first treatment policies are put in place, the worst-off individuals are those with intermediate centrality (everyone's risk reduces somewhat). 

Looking at expected total number of sick days in the entire population, the different central-first policies are all similar (No treatments: 58.01, Eigen-centrality: 25.56, Degree-centrality: 25.61; Betweenness-centrality: 25.17), but the individual burdens are different (Figure~\ref{fig:karate_club_centrality}). The Deep Q-network (DQN) did not succeed in learning a policy that outperformed centrality heuristics, suggesting that a different network structure or training approach should be tried.

The Karate club graph naturally splits into two communities based on how members of the club sided in a particular conflict that led to the dissolution of the club (for more history on this graph, see~\citet{zachary1977information}). Figure~\ref{fig:karate_club_community} shows how the disease burden is spread over the two communities (details on the community detection procedure is in Appendix~\ref{app:community_dection_karate}). On average, community 1 tends to fare a little bit better than community 2. This is even true with natural disease progression, suggesting that community 1 is somehow structured in a way that is less conducive to disease spread. 
The DQN favors community 1 in an extreme way, but since it's performance is generally poor, its hard to draw strong conclusions from that.

\section{Conclusion}
This is work in progress, but we believe that the precision contagion treatment problem, which we intend to contribute to the collection of simulated environments in the open-source {\tt ML-fairness-gym}~\citep{fairnessgym}, contains interesting questions of how to find fair allocation policies when individuals are highly intertwined. 
It emphasizes the importance of {\em fairness in  outcomes} rather than a narrower goal of fairness in e.g., classification accuracy (there is no classification or state estimation task in this problem -- just planning).
It also highlights how individuals with ``intermediate risk'' can end up being most at risk worst when policies are put in place to help the most at risk.
Despite the idealized nature of the problem which assumes fully observed graph structures and highly precise treatment delivery, we hope that an increased understanding of the fairness implications of the underlying optimization problem with network effects can inform work in healthcare more broadly.

\bibliographystyle{plainnat}
\bibliography{references}

\newpage
\appendix
\section{Appendix}

This appendix contains a description of the methodology used to solve for treatment strategies in chain graphs, the methodology used to train a Q-learner on the Karate Club network, and a brief description of community detection techniques that were used on the Karate Club network.

\subsection{Solving for treatment strategies in the chain graph via linear program}
\label{app:chain_lp}

\begin{figure}
    \centering
    \includegraphics[width=0.33\columnwidth]{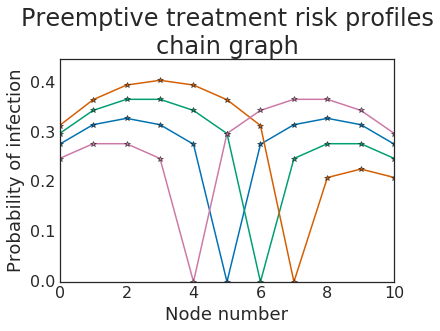}
    \caption{Some risk profiles in a chain graph with 11 nodes for different preemptive treatment strategies. Treating node $i$ creates zero risk for individual $i$ and lowers the risk for neighbors as well.}
    \label{fig:chain_graph_bases}
\end{figure}

\subsubsection{Preemptive treatments}
In the preemptive setting, with only a single treatment, there are a limited number of possible interventions, corresponding to treating each of the members of the graph. 
Each of those interventions has a corresponding risk profile (Figure~\ref{fig:chain_graph_bases}).

To find the strategy that minimizes average disease burden, it is sufficient to simply enumerate them and find the lowest.

To see if there is a distribution over strategies that gives a uniform risk profile, we try to find a convex combination of risk profiles that minimizes average risk and is subject to the constraints that all of the risks are equal. These are all linear constraints, so the problem is a linear program. We use scipy's {\tt optimize.linprog} function as a solver.

\subsubsection{1-step precision treatments}
We also consider precision treatments after 1 step - i.e., waiting to see which patient is infected and responding with a treatment at that point.

For a chain graph with N members, we are now looking for N different reactive allocations, with each allocation being a convex combination of treatment risk profiles, similar to above, but with the initial patient already infected.

This problem has a similar form to above, but with N convex combination constraints instead of 1, and is also a linear program.

Waiting for longer than one step before acting was not considered, though could be a reasonable strategy as well.

\subsection{Training a DQN network on the karate-club graph}
\label{app:training_dqn}

A DQN agent was trained using dopamine's experiment runner.
The reward function at each step was the negative number of newly sick nodes.

The observation vector at each step was one-hot encoded health states. 

The graph structure was not passed to the agent explicitly, but it was given the chance to learn by observing the simulations.

Hyperparameters of the agent were optimized using a black-box hyperparameter tuning system.
%similar to the one described in~\citet{vizier}. 

The range of settings explored were:
\begin{itemize}
    \item gamma: [0, 0.99]
    \item hidden layer size: [5, 500]
    \item learning rate: [$10^{-4}$, $10^0$]
    \item num iterations: [$10^2$, $10^5$]
\end{itemize}
The agent's stack size was set to 1; max steps per episode was 20, and the training steps per iteration was 500. All other arguments were left at default.

The best settings found by the hyperparameter tuner after 150 trials were:
\begin{itemize}
    \item gamma: 0.70
    \item hidden layer size: 106
    \item learning rate: 0.95
    \item num iterations: 2543
\end{itemize}

\subsection{Community detection in the karate-club graph}
\label{app:community_dection_karate}
\begin{figure}
    \centering
    \includegraphics[width=0.5\columnwidth]{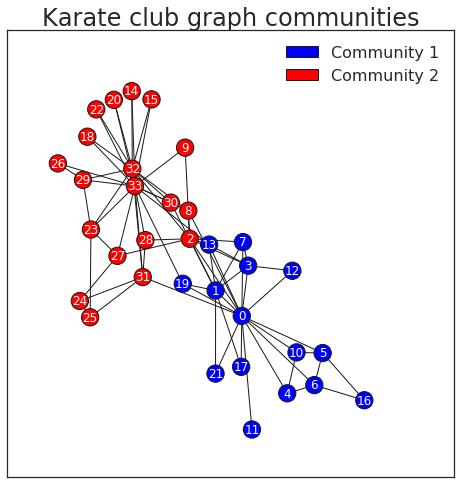}
    \caption{Partitioning of the Karate club graph into 2 communities using the Girvan-Newman algorithm}
    \label{fig:karate_partition}
\end{figure}

We use networkx's implementation of the Girvan Newman algorithm for community detection and use the highest level in the hierarchy, creating two communities shown in Figure~\ref{fig:karate_partition}. 
\end{document}